\documentclass[conference]{IEEEtran}
\usepackage{geometry}
\usepackage{amsmath, amsfonts, amssymb}
\usepackage{graphicx}
\usepackage[breaklinks]{hyperref}
\usepackage{url}
\usepackage{breakurl}
\usepackage{cite}
\usepackage{dirtytalk}
\usepackage{booktabs}
\usepackage{array}
\usepackage{float}

\expandafter\def\expandafter\UrlBreaks\expandafter{\UrlBreaks
    \do\a\do\b\do\c\do\d\do\e\do\f\do\g\do\h\do\i\do\j%
    \do\k\do\l\do\m\do\n\do\o\do\p\do\q\do\r\do\s\do\t%
    \do\u\do\v\do\w\do\x\do\y\do\z\do\A\do\B\do\C\do\D%
    \do\E\do\F\do\G\do\H\do\I\do\J\do\K\do\L\do\M\do\N%
    \do\O\do\P\do\Q\do\R\do\S\do\T\do\U\do\V\do\W\do\X%
    \do\Y\do\Z\do\/\do-}

\title{Dark Deceptions in DHCP: Dismantling Network Defenses}

\author{
    \IEEEauthorblockN{Robert Dilworth}
    \IEEEauthorblockA{
        {Department of Computer Science and Engineering} \\
        {Mississippi State University} \\
        {Starkville, Mississippi, USA} \\
        {\href{https://orcid.org/0009-0005-5497-9810}{rkd103@msstate.edu}}
    }
}

\begin{document}

\maketitle

\begin{abstract}
    This paper explores vulnerabilities in the Dynamic Host Configuration Protocol (DHCP) and their implications on the Confidentiality, Integrity, and Availability (CIA) Triad. Through an analysis of various attacks, including DHCP Starvation, Rogue DHCP Servers, Replay Attacks, and TunnelVision exploits, the paper provides a taxonomic classification of threats, assesses risks, and proposes appropriate controls. The discussion also highlights the dangers of VPN decloaking through DHCP exploits and underscores the importance of safeguarding network infrastructures. By bringing awareness to the TunnelVision exploit, this paper aims to mitigate risks associated with these prevalent vulnerabilities.
\end{abstract}


\section{Introduction}
\label{sec:Introduction}

    Dynamic Host Configuration Protocol (DHCP) plays a pivotal role in modern network management by automating the allocation of IP addresses to devices. However, its simplicity also makes it a prime target for attackers seeking to disrupt network operations. This paper will explore the vulnerabilities inherent in DHCP, focusing specifically on the TunnelVision exploit (CVE-2024-3661) and its profound implications on VPN security. By detailing these attacks and classifying their risks, we aim to highlight how they affect the Confidentiality, Integrity, and Availability (CIA) Triad, a fundamental framework for assessing the security of information systems.

    With a clear understanding of the significance and vulnerabilities of DHCP, it is crucial to delve deeper into the background of the protocol itself, particularly how it interacts with the core principles of network security.

    \subsection{Thesis}
    \label{subsec:Thesis}

        Despite its widespread adoption, DHCP remains vulnerable to several forms of attack, some of which can undermine even the most secure systems. The TunnelVision exploit, for instance, demonstrates how DHCP can be weaponized to bypass VPN security protocols, exposing sensitive data. This paper aims to provide an in-depth analysis of such vulnerabilities, focusing on their potential to compromise the CIA Triad and offering practical solutions for mitigation.

        Building on this perspective, it is important to explore why these vulnerabilities have far-reaching implications for network security, especially in relation to the Confidentiality, Integrity, and Availability of systems.

    \subsection{Justification}
    \label{subsec:Justification}
        
        The existence of DHCP vulnerabilities poses significant risks to network security. Attacks targeting DHCP often occur covertly, leaving networks discombobulated and users precariously exposed. By highlighting these vulnerabilities, this paper aims to obviate their impacts and promulgate awareness.

        Given the gravity of these threats, it becomes evident that a comprehensive analysis of specific DHCP attack types is necessary to fully grasp the risks and the corresponding solutions.

    \subsection{Contributions}
    \label{subsec:Contributions}
    
        The chief contribution of this paper lies in raising awareness about the TunnelVision DHCP exploit and its implications for VPN security. Additionally, the paper offers a granular classification of DHCP attacks, analyzes their risks, and suggests corrective, detective, and mitigative controls.

    \subsection{Paper Outline}
    \label{subsec:Paper_Outline}

        This paper is organized into several sections to systematically address the vulnerabilities within the Dynamic Host Configuration Protocol (DHCP) and their implications on network security. Following this introduction (Section \ref{sec:Introduction}), Section \ref{sec:Background} provides sufficient background knowledge on the DHCP protocol and its interaction with the Confidentiality, Integrity, and Availability (CIA) Triad. In Section  \ref{sec:Taxonomic_Classification_of_DHCP_Attacks}, we introduce a taxonomic classification of common DHCP attacks, including DHCP Starvation, Rogue DHCP Servers, Replay Attacks, and VPN Decloaking through the TunnelVision exploit. These attacks are then discussed in detail in Section  \ref{sec:Discussion_of_DHCP_Attacks}, where we examine their impact on the CIA Triad and explore corrective, detective, and mitigative controls for each. The paper delves deeper into the specific risks posed by the TunnelVision exploit and its effect on VPN security in Section  \ref{sec:The_TunnelVision_Exploit}. As part of ongoing research and future directions, Section \ref{sec:Future_Work} explores the application of PU/NU learning techniques to the detection of TunnelVision attacks, offering new perspectives on automated detection and mitigation strategies for DHCP-related security threats. Finally, in the conclusion (Section  \ref{sec:Conclusion}), the paper emphasizes the critical need for heightened vigilance in securing DHCP and maintaining robust defenses against these attacks.

        Now that the paper's structure is laid out, we can turn to the foundational concepts surrounding DHCP to ensure a comprehensive understanding of the protocol before diving into the specific vulnerabilities.
    
    \subsection{Acronyms Utilized}
    \label{subsec:Acronyms_Utilized}

        \begin{itemize}
            \item ARP (\textit{Address Resolution Protocol}): Maps IP addresses to MAC addresses.
            \item CIA (\textit{Confidentiality, Integrity, Availability})
            \item CIDR (\textit{Classless Inter-Domain Routing}): IP address allocation without fixed classes.
            \item CVE (\textit{Common Vulnerabilities and Exposures}): A database of publicly disclosed vulnerabilities.
            \item DHCP (\textit{Dynamic Host Configuration Protocol}): Protocol for dynamically assigning IP addresses.
            \item DHCPv6 (\textit{Dynamic Host Configuration Protocol for IPv6}): DHCP for IPv6 addressing.
            \item DNS (\textit{Domain Name System}): Translates domain names to IP addresses.
            \item HTTPS (\textit{Hypertext Transfer Protocol Secure}): Secure version of HTTP.
            \item IPv6 (\textit{Internet Protocol version 6}): The most recent version of the Internet Protocol.
            \item MITM (\textit{Man-in-the-Middle}): An attacker intercepts and alters communication.
            \item PPTP (\textit{Point-to-Point Tunneling Protocol}): A protocol for creating VPNs.
            \item VPN (\textit{Virtual Private Network}): A secure private network over the internet.
            \item WebRTC (\textit{Web Real-Time Communication}): Real-time communication over the web.
        \end{itemize}

        With these terms clarified, we can proceed to examine the fundamental workings of DHCP, focusing on how it facilitates network configuration and the potential risks it introduces in the absence of strong security features.

    \subsection{Tables Used Throughout the Paper}
    \label{subsec:Tables}
        
        Several tables have been incorporated throughout the paper to summarize key concepts and provide detailed insights into the various DHCP attacks discussed.
        
        \begin{itemize}
            \item \textbf{Table \ref{tab:DHCP_Starvation_Attack}}: This table provides an overview of the DHCP Starvation Attack, detailing its definition, impact on the CIA Triad, and the recommended controls to mitigate its effects.
            \item \textbf{Table \ref{tab:Rogue_DHCP_Servers}}: This table outlines the Rogue DHCP Server Attack, explaining how malicious DHCP servers can compromise network configurations and security. It also discusses the attack’s implications on Confidentiality, Integrity, and Availability, alongside suggested mitigation strategies.
            \item \textbf{Table \ref{tab:Replay_Attacks}}: This table highlights the Replay Attack, focusing on how attackers can replay intercepted DHCP packets to impersonate clients or servers. The table also categorizes the impact of the attack on the CIA Triad and offers recommendations for countermeasures.
            \item \textbf{Table \ref{tab:VPN_Decloaking}}: This table discusses the TunnelVision (VPN Decloaking) Attack, where DHCP mechanisms are exploited to bypass VPN security. The table addresses the attack’s impact on the CIA Triad and the necessary controls to safeguard VPN traffic from such vulnerabilities.
        \end{itemize}

\section{Background}
\label{sec:Background}

    \subsection{DHCP Fundamentals}
    \label{subsec:DHCP_Fundamentals}

        DHCP simplifies network management by automating IP address allocation, configuration, and management. The protocol operates through a series of message exchanges:

        \begin{itemize}
            \item \texttt{DHCPDISCOVER}: Broadcasted by a client to locate available DHCP servers.
            \item \texttt{DHCPOFFER}: Sent by a server offering an IP address lease.
            \item \texttt{DHCPREQUEST}: Used by the client to accept the offer.
            \item \texttt{DHCPACK}: Sent by the server to confirm the lease and finalize configuration.
        \end{itemize}

        This \textit{convenience} comes at a cost, as the lack of inherent security features makes DHCP vulnerable to various attacks, including DHCP Starvation and Rogue DHCP server attacks.

    \subsection{The CIA Triad}
    \label{subsec:The_CIA_Triad}
        
        The CIA Triad encapsulates the core principles of cybersecurity: Confidentiality, Integrity, and Availability.
        
        \begin{itemize}
            \item \textbf{Confidentiality}: Ensuring information is accessible only to authorized users.
            \item \textbf{Integrity}: Maintaining the accuracy and trustworthiness of data.
            \item \textbf{Availability}: Guaranteeing information is accessible when required. 
        \end{itemize}

        The three types of \textit{controls}--or measures used to manage risk--that can be applied to address the above principles are:

        \begin{itemize}
            \item \textbf{Corrective}: Actions taken to restore systems after an incident.
            \item \textbf{Detective}: Measures to identify and alert on security issues.
            \item \textbf{Mitigative}: Strategies to reduce the impact of potential threats.
        \end{itemize}
        
        DHCP attacks often undermine these principles, necessitating robust controls to safeguard network assets.

    Having established a foundational understanding of DHCP and the CIA Triad, it is now necessary to categorize and discuss the various attack vectors that exploit these vulnerabilities.

\section{Taxonomic Classification of DHCP Attacks}
\label{sec:Taxonomic_Classification_of_DHCP_Attacks}

    Attacks exploiting DHCP vulnerabilities can be classified as follows:

    \begin{enumerate}
        \item \textbf{DHCP Starvation Attacks:} Exhaust IP address pools to deny legitimate client access \cite{7h30th3r0n32024, Mikhailov2024, AbdulGhaffar2023}.
        \item \textbf{Rogue DHCP Servers:} Deploy malicious servers to deliver compromised configurations \cite{7h30th3r0n32024, Mikhailov2024, AbdulGhaffar2023}.
        \item \textbf{Replay Attacks:} Resend intercepted DHCP packets to impersonate devices \cite{AbdulGhaffar2023}.
        \item \textbf{VPN Decloaking (TunnelVision):} Exploit DHCP mechanisms to bypass VPN security \cite{Moratti2024}.
    \end{enumerate}

    With this classification framework in place, we can now examine each attack in detail, starting with the DHCP Starvation Attack, and explore its impacts on network functionality and security.

\section{Discussion of DHCP Attacks}
\label{sec:Discussion_of_DHCP_Attacks}

    \subsection{DHCP Starvation Attack}
    \label{subsec:DHCP_Starvation_Attack}

        This attack, indicative of a denial-of-service (DoS) approach, inundates the DHCP server with spoofed \texttt{DHCPDISCOVER} requests. By exhausting the server’s IP pool, legitimate clients are left without network access. See Table \ref{tab:DHCP_Starvation_Attack}. 

        \begin{table*}[h!]
            \centering
            \caption{Overview of DHCP Starvation Attack}
            \label{tab:DHCP_Starvation_Attack}
            \begin{tabular}{@{}>{\raggedright}p{3cm} p{10cm}@{}}
                \toprule
                \textbf{Category} & \textbf{Details} \\ 
                \midrule
                \textbf{Definition} & Attackers exhaust the server's IP address pool by flooding it with \texttt{DHCPDISCOVER} requests using spoofed MAC addresses. This prevents legitimate users from obtaining IP addresses, causing a Denial-of-Service (DoS). \\ 
                \midrule
                \textbf{Impact on CIA Triad} & 
                \textbf{Confidentiality:} Limited impact unless combined with other attacks. \\
                & \textbf{Integrity:} Spoofed MAC addresses undermine trust in DHCP assignments. \\
                & \textbf{Availability:} Severely impacted as legitimate clients cannot obtain IP addresses. \\ 
                \midrule
                \textbf{Controls} & 
                \textbf{Corrective:} Configure IP address thresholds. \\
                & \textbf{Detective:} Monitor unusual DHCP request patterns. \\
                & \textbf{Mitigative:} Employ DHCP snooping on switches. \\ 
                \bottomrule
            \end{tabular}
        \end{table*}

    \subsection{Rogue DHCP Servers}
    \label{subsec:Rogue_DHCP_Servers}

        In this attack, an attacker sets up a malicious DHCP server, siphoning legitimate client traffic through compromised gateways. See Table \ref{tab:Rogue_DHCP_Servers}.

        \begin{table*}[h!]
            \centering
            \caption{Overview of Rogue DHCP Servers}
            \label{tab:Rogue_DHCP_Servers}
            \begin{tabular}{@{}>{\raggedright}p{3cm} p{10cm}@{}}
                \toprule
                \textbf{Category} & \textbf{Details} \\ 
                \midrule
                \textbf{Definition} & A rogue server issues malicious configurations, such as redirecting traffic to attacker-controlled gateways or DNS servers. This enables man-in-the-middle (MITM) attacks, phishing, and traffic interception. \\ 
                \midrule
                \textbf{Impact on CIA Triad} & 
                \textbf{Confidentiality:} Traffic interception enables data theft. \\
                & \textbf{Integrity:} Malicious configurations distort network routing. \\
                & \textbf{Availability:} Misconfigured clients experience service disruptions. \\ 
                \midrule
                \textbf{Controls} & 
                \textbf{Corrective:} Validate DHCP server configurations. \\
                & \textbf{Detective:} Use network anomaly detection tools. \\
                & \textbf{Mitigative:} Implement ARP binding and DHCP snooping. \\ 
                \bottomrule
            \end{tabular}
        \end{table*}

    \subsection{Replay Attacks}
    \label{subsec:Replay_Attacks}

        Such an attack capitalizes on vulnerabilities in the DHCP protocol where session information lacks nonces or unique identifiers. By capturing and retransmitting legitimate packets, attackers can impersonate clients or servers, leading to unauthorized access, network disruption, and data exposure. These attacks are particularly insidious because they often go undetected during routine network operations. See Table \ref{tab:Replay_Attacks}.

        \begin{table*}[h!]
            \centering
            \caption{Overview of Replay Attacks}
            \label{tab:Replay_Attacks}
            \begin{tabular}{@{}>{\raggedright}p{3cm} p{10cm}@{}}
                \toprule
                \textbf{Category} & \textbf{Details} \\ 
                \midrule
                \textbf{Definition} & Replay attacks exploit the absence of nonces in DHCP communications. An attacker captures legitimate DHCP packets and retransmits them to masquerade as a valid client or server. This allows unauthorized network access and potential traffic manipulation. \\ 
                \midrule
                \textbf{Impact on CIA Triad} & 
                \textbf{Confidentiality:} Unauthorized access exposes sensitive information during packet interception. \\
                & \textbf{Integrity:} Manipulation of replayed packets disrupts routing and can inject malicious data. \\
                & \textbf{Availability:} Repeated replays can overwhelm servers, causing delays or service unavailability. \\ 
                \midrule
                \textbf{Controls} & 
                \textbf{Corrective:} Authenticate DHCP communications using session-based nonces or timestamps. \\
                & \textbf{Detective:} Monitor network traffic for duplicate DHCP requests using intrusion detection systems (IDS). \\
                & \textbf{Mitigative:} Implement DHCP authentication options such as Option 82 or adopt secure extensions like DHCPv6. \\ 
                \bottomrule
            \end{tabular}
        \end{table*}

    \subsection{VPN Decloaking (TunnelVision)}
    \label{subsec:VPN_Decloaking}

        TunnelVision exploits DHCP to redirect VPN traffic through attacker-controlled routes, thwarting encryption and anonymity. See Table \ref{tab:VPN_Decloaking}.

        \begin{table*}[h!]
            \centering
            \caption{Overview of VPN Decloaking (TunnelVision)}
            \label{tab:VPN_Decloaking}
            \begin{tabular}{@{}>{\raggedright}p{3cm} p{10cm}@{}}
                \toprule
                \textbf{Category} & \textbf{Details} \\ 
                \midrule
                \textbf{Definition} & Moratti et al. \cite{Moratti2024} detail how attackers use DHCP option 121 to manipulate VPN traffic. By configuring routes more specific than \texttt{/0} CIDR ranges, attackers bypass encrypted tunnels, forcing traffic through rogue gateways. This exposes sensitive data to interception and modification. \\ 
                \midrule
                \textbf{Impact on CIA Triad} & 
                \textbf{Confidentiality:} Exposes VPN-encrypted traffic. \\
                & \textbf{Integrity:} Allows malicious route manipulation. \\
                & \textbf{Availability:} Potential disruptions in VPN connectivity. \\ 
                \midrule
                \textbf{Controls} & 
                \textbf{Corrective:} Harden DHCP client configurations. \\
                & \textbf{Detective:} Monitor DHCP traffic for anomalies. \\
                & \textbf{Mitigative:} Use static routes where feasible. \\ 
                \bottomrule
            \end{tabular}
        \end{table*}

    Following this discussion of individual attack types, it is critical to focus specifically on the TunnelVision exploit. This attack exemplifies a particularly sophisticated manipulation of DHCP, bypassing the security measures of VPNs.

\section{The TunnelVision Exploit: A Frightening Prospect for VPN Security}
\label{sec:The_TunnelVision_Exploit}

    The TunnelVision exploit, as detailed by \cite{Moratti2024}, is a glaring reminder of the fragile boundary between security and exposure in our hyperconnected world. By leveraging DHCP vulnerabilities, this attack circumvents the encrypted tunnels that VPNs rely on, exposing user traffic to a rogue DHCP server under the attacker’s control. It is a sobering development in network security, signaling a seismic shift in how adversaries can undermine even the most trusted privacy measures.  

    \subsection{Breaking the Tunnel, Breaching the Trust}
    \label{subsec:Breaking_the_Tunnel}

        For years, Virtual Private Networks (VPNs) have been marketed as the cornerstone of privacy, promising encrypted communication and anonymity against prying eyes. However, as highlighted by \cite{Abbas2023}, these promises are often contingent on precise configurations. The moment users neglect to verify server certificates or disable IPv6, they unknowingly expose themselves to a gamut of threats, including DNS leaks and MITM attacks. The TunnelVision exploit takes this one step further, not merely bypassing misconfigurations but fundamentally dismantling VPN integrity through DHCP manipulation.  

        In the TunnelVision attack, the adversary first sets up a rogue DHCP server, using techniques such as DHCP Starvation or ARP spoofing to ensure their responses reach the targeted client first. By exploiting DHCP option 121, the attacker injects routes into the client’s routing table, effectively rerouting VPN-bound traffic back into the attacker’s network. This results in catastrophic breaches of Confidentiality, Integrity, and Availability--pillars of the CIA Triad that underpin modern security frameworks.  

    \subsection{A Seed of Paranoia for Privacy Advocates}
    \label{subsec:A_Seed_of_Paranoia}

        The discovery of this exploit sows the seeds of paranoia in anyone concerned with their digital privacy. When a VPN--considered the last bastion of security--fails, users are left vulnerable to their device’s baseline encryption protocols, such as HTTPS. If these protections falter, the implications are stark: passwords, financial transactions, and private communications are exposed in plain text. To put it bluntly, the failure of VPN security renders users irrevocably \say{pwned.}  

        \cite{Moratti2024} emphasizes that this attack is not an esoteric proof-of-concept but a feasible strategy that exploits standard DHCP functionality. The deliberate design of option 121 allows attackers to selectively divert traffic, bypassing VPN tunnels with surgical precision. Combined with the findings of \cite{Abbas2023}, it is clear that VPNs are not invulnerable. Their security depends on meticulous configuration and constant vigilance, as missteps can lead to devastating consequences, including de-anonymization and data theft.  

    \subsection{Implications and Countermeasures}
    \label{subsec:Implications_and_Countermeasures}

        The TunnelVision exploit starkly highlights the broader vulnerabilities in network protocols. Its success demonstrates that VPNs alone are insufficient to ensure privacy. As \cite{Abbas2023} notes, attackers increasingly exploit VPN misconfigurations and ancillary leaks like DNS or WebRTC to expose sensitive information. TunnelVision capitalizes on these weaknesses, effectively using DHCP as a trojan horse to dismantle encrypted communication channels.  

        Addressing this requires a multi-layered approach. Users must adopt VPN solutions that rigorously enforce route integrity, including safeguards against arbitrary route injection. System administrators can mitigate such risks by implementing DHCP snooping and using MAC binding to prevent unauthorized DHCP servers from operating within the network. Additionally, promoting best practices such as routinely updating VPN software, enabling necessary protocols like kill-switches \cite{Mullvad2024,ProtonVPN2024}, and enabling DNS leak protection can close potential attack vectors.  

    \subsection{A Bell of Awakening for Secure Communications}
    \label{subsec:A_Bell_of_Awakening}

        The TunnelVision exploit is more than just a CVE curiosity--it is a wake-up call. It underscores the urgent need to revisit the foundational assumptions of network security. VPNs, long hailed as the panacea for privacy concerns, are now revealed to be fragile constructs susceptible to protocol-level manipulation. Without a concerted effort to address these vulnerabilities, the promise of secure communication will remain an illusion.  

        The TunnelVision exploit highlights the odious foibles of relying solely on VPNs for privacy. Its deployment underscores the importance of understanding DHCP’s vulnerabilities and biding vigilance in network design.

\section{Future Work}
\label{sec:Future_Work}

    To expand the research on TunnelVision and its detection mechanisms, we introduce the concept of PU/NU (Positive Unlabeled/Negative Unlabeled) Learning. The TunnelVision exploit can be framed as a machine learning problem, where the challenge is detecting malicious DHCP lease events amidst a large volume of benign traffic.

    \subsection{TunnelVision (CVE-2024-3661) as a PU/NU Learning Problem}
    
        \subsubsection{Overview of TunnelVision}
        
            TunnelVision (CVE-2024-3661) is a recently discovered network security vulnerability that allows attackers to bypass VPN encapsulation by exploiting DHCP (Dynamic Host Configuration Protocol) functionality. Specifically, the exploit leverages DHCP option 121 to inject malicious routing rules into the victim’s network stack, leading to the decloaking of the victim’s VPN. This attack causes some of the victim's traffic to bypass the encrypted VPN tunnel, making sensitive data vulnerable to interception.
            
            An attacker can execute the attack by running a rogue DHCP server on the same network as the target. By triggering DHCP lease renewal or expiration events, the attacker can push malicious routing entries that take precedence over the VPN’s default routes. This makes detection challenging because DHCP client logs typically contain benign lease expiration and installation events, masking the malicious behavior within seemingly normal traffic. Detecting such anomalies represents an opportunity for applying PU/NU learning.
        
        \subsubsection{Problem Formulation}
        
            The core challenge in detecting TunnelVision exploits lies in identifying malicious DHCP lease events (positive examples) among a large number of benign DHCP logs (unlabeled or negative events). DHCP lease logs usually contain high volumes of legitimate traffic, with malicious events occurring rarely. Therefore, this problem can be formulated as a PU/NU learning task, where the system must distinguish between benign and malicious lease events with limited labeled data.
        
        \subsubsection{Mathematical Formulation}
        
            Let:
            
            \begin{equation}
                L = \{l_1, l_2, \dots, l_n\}
            \end{equation}
            
            represent the set of all DHCP lease logs, where \(l_i \in L\) is an individual lease event related to DHCP lease expiration or installation.
            
            \begin{equation}
                P \subseteq L
            \end{equation}
            
            is the set of known malicious lease events related to TunnelVision (positive examples) and
            
            \begin{equation}
                U = L \setminus P
            \end{equation}
            
            is the set of unlabeled logs, which may contain both benign and malicious events.
            
            The goal is to build a classifier \(f: L \to \{0, 1\}\) where:
            
            \begin{equation}
                f(l_i) = 1
            \end{equation}
            
            indicates that the lease event \(l_i\) is malicious, and
            
            \begin{equation}
                f(l_i) = 0
            \end{equation}
            
            indicates that the lease event \(l_i\) is benign.
            
            The risk function \(R(f)\) for this PU learning problem is defined as:
            
            \begin{equation}
                R(f) = \mathbb{E}_{x \in P} [l(f(x), 1)] + \mathbb{E}_{x \in U} [l(f(x), -1)],
            \end{equation}
            
            where \(l(f(x), y)\) is a loss function that measures the difference between predicted and true labels.
            
        \subsubsection{Features of the Logs}
        
            Several features of DHCP logs can help distinguish benign events from malicious ones:
            
            \begin{itemize}
                \item \textbf{Lease Expiration Time}: A shorter lease expiration time may indicate malicious activity, as attackers could exploit this to force frequent lease renewals.
                \item \textbf{Option 121 Usage}: The presence and frequency of DHCP option 121 are key indicators of TunnelVision attempts.
                \item \textbf{Routing Table Changes}: Logs indicating changes to the routing table (e.g., the insertion of more specific routes via option 121) may suggest malicious behavior.
            \end{itemize}
        
        \subsubsection{Adapting NU Learning for Benign Logs}
        
            If a set of benign DHCP lease events can be reliably labeled, the problem can be reformulated as an NU learning task. In this case, let \(N \subseteq L\) represent the set of known benign lease events (negative examples), and \(U = L \setminus N\) is the remaining set of unlabeled events. The objective in NU learning is to detect positive (malicious) samples from the unlabeled set.
        
        \subsubsection{Expected Results and Implementation}
        
            By applying PU and NU learning techniques, we expect to achieve the following outcomes:
            
            \begin{itemize}
                \item \textbf{Improved Detection of TunnelVision Attacks}: The model can effectively flag malicious lease events, enabling early detection and preemptive action by network administrators.
                \item \textbf{Scalable Log Processing}: The PU/NU learning approach allows for the analysis of large volumes of DHCP lease logs, even when labeled data is scarce and malicious events are infrequent.
                \item \textbf{Early Detection}: By focusing on subtle anomalies in DHCP lease behavior, such as frequent lease renewals or the use of Option 121, the system can detect attacks at their early stages.
            \end{itemize}
            
            Performance will be evaluated using Precision, Recall, and \(F_{1}\)-score to balance between detecting true positives (malicious events) and minimizing false positives.
    
\section{Conclusion}
\label{sec:Conclusion}

    This paper underscores the ascendancy of DHCP as both a critical and precarious element in network management. By ascribing attention to its vulnerabilities, particularly the TunnelVision exploit, we hope to squelch the risks these attacks pose. 

    Rather, DHCP's simplicity and ubiquity make it indispensable but also a high-value target for attackers. Understanding its vulnerabilities and the implications of attacks like TunnelVision is essential for designing secure networks.

\bibliographystyle{IEEEtran}
\bibliography{references}

\begin{thebibliography}{1}
\providecommand{\url}[1]{#1}
\csname url@samestyle\endcsname
\providecommand{\newblock}{\relax}
\providecommand{\bibinfo}[2]{#2}
\providecommand{\BIBentrySTDinterwordspacing}{\spaceskip=0pt\relax}
\providecommand{\BIBentryALTinterwordstretchfactor}{4}
\providecommand{\BIBentryALTinterwordspacing}{\spaceskip=\fontdimen2\font plus
\BIBentryALTinterwordstretchfactor\fontdimen3\font minus \fontdimen4\font\relax}
\providecommand{\BIBforeignlanguage}[2]{{%
\expandafter\ifx\csname l@#1\endcsname\relax
\typeout{** WARNING: IEEEtran.bst: No hyphenation pattern has been}%
\typeout{** loaded for the language `#1'. Using the pattern for}%
\typeout{** the default language instead.}%
\else
\language=\csname l@#1\endcsname
\fi
#2}}
\providecommand{\BIBdecl}{\relax}
\BIBdecl

\bibitem{7h30th3r0n32024}
\BIBentryALTinterwordspacing
h30th3r0n3, ``Evil-m5project v1.3.6 - network hijacking,'' \emph{GitHub}. [Online]. Available: \url{https://github.com/7h30th3r0n3/Evil-M5Core2}
\BIBentrySTDinterwordspacing

\bibitem{Mikhailov2024}
\BIBentryALTinterwordspacing
A.~Mikhailov, ``Attacks on the dhcp protocol: Dhcp starvation, dhcp spoofing, and protection against these techniques,'' \emph{HackMag}. [Online]. Available: \url{https://hackmag.com/security/dhcp-hacking/}
\BIBentrySTDinterwordspacing

\bibitem{AbdulGhaffar2023}
\BIBentryALTinterwordspacing
A.~AbdulGhaffar, S.~K. Paul, and A.~Matrawy, ``An analysis of dhcp vulnerabilities, attacks, and countermeasures,'' in \emph{2023 Biennial Symposium on Communications (BSC)}.\hskip 1em plus 0.5em minus 0.4em\relax IEEE, 7 2023, pp. 119--124. [Online]. Available: \url{https://ieeexplore.ieee.org/document/10201458}
\BIBentrySTDinterwordspacing

\bibitem{Moratti2024}
\BIBentryALTinterwordspacing
L.~Moratti and D.~Cronce, ``Tunnelvision (cve-2024-3661): How attackers can decloak routing-based vpns for a total vpn leak,'' \emph{Leviathan Security Group}, 5 2024. [Online]. Available: \url{https://www.leviathansecurity.com/blog/tunnelvision}
\BIBentrySTDinterwordspacing

\bibitem{Abbas2023}
\BIBentryALTinterwordspacing
H.~Abbas, N.~Emmanuel, M.~F. Amjad, T.~Yaqoob, M.~Atiquzzaman, Z.~Iqbal, N.~Shafqat, W.~B. Shahid, A.~Tanveer, and U.~Ashfaq, ``Security assessment and evaluation of vpns: A comprehensive survey,'' \emph{ACM Computing Surveys}, vol.~55, pp. 1--47, 12 2023. [Online]. Available: \url{https://dl.acm.org/doi/10.1145/3579162}
\BIBentrySTDinterwordspacing

\bibitem{Mullvad2024}
\BIBentryALTinterwordspacing
Mullvad, ``Evaluating the impact of tunnelvision,'' \emph{Mullvad VPN AB}, 5 2024. [Online]. Available: \url{https://mullvad.net/en/blog/evaluating-the-impact-of-tunnelvision}
\BIBentrySTDinterwordspacing

\bibitem{ProtonVPN2024}
\BIBentryALTinterwordspacing
ProtonVPN, ``Tunnelvision and proton vpn,'' \emph{Proton AG}. [Online]. Available: \url{https://protonvpn.com/support/tunnelvision/}
\BIBentrySTDinterwordspacing

\end{thebibliography}

\end{document}